\documentclass[12pt,preprint]{aastex}
\usepackage{lineno}
\shorttitle{Pulsar Magnetosphere Dynamics}

\begin{document}

\title{Pulsar Magnetosphere Dynamics}

\author{K.H. Tsui}
\affil{Instituto de F\'{i}sica - Universidade Federal Fluminense
\\Campus da Praia Vermelha, Av. General Milton Tavares de Souza s/n
\\Gragoat\'{a}, 24.210-346, Niter\'{o}i, Rio de Janeiro, Brasil.}

%\\author{K.H. Tsui$^{\dag}$, J.A. Arteaga$^{\dag \dag}$, A. Serbeto$^{\dag}$}
%\\affil{$^\dag$ Instituto de F\'{i}sica - Universidade Federal Fluminense
%\\\Campus da Praia Vermelha, Av. General Milton Tavares de Souza s/n
%\\\Gragoat\'{a}, 24.210-346, Niter\'{o}i, Rio de Janeiro, Brasil.
%\\\
%\$^{\dag \dag}$ ITR Centro Sur - Universidad Tecnol\'ogica  del Uruguay, Durazno  97000, Uruguay.}

\email{tsui$@$if.uff.br}
\pagestyle{myheadings}
\baselineskip 18pt
	
\begin{abstract}

With the neutron star rotating under a stationary magnetic field
 generating unipolar induction,
 charges are driven to the pulsar surface according to their signs,
 and are uploaded to the magnetosphere along the magnetic field lines.
 In the presence of an electron-positron plasma,
 the pulsar magnetosphere is represented
 by a unique magnetohydrodynamic (MHD) plasma.
 The pulsar equation of force-free equilibrium
 is solved analytically for two configurations.
 The first one is a light cylinder guided jet-like open magnetosphere,
 and the second one is a closed magnetosphere within the light cylinder.
 The Goldreich Julian condition for the stability
 of the closed magnetosphere (dead zone) is reconsidered
 which indicates transition of the closed magnetosphere to the open one
 as the electron-positron plasma density builds up.
 This suggests that the pulsar period
 could be the result of magnetosphere dynamics
 rather than the pulsar rotation.

\end{abstract}

\maketitle

\newpage
\section{Pulsar Electron-Positron Magnetohydrodynamic Plasma}

The basic structure of the current pulsar magnetosphere
 has been set over the decades by many distinguished pioneers
 \citep{goldreich1969, sturrock1971, michel1973, scharlemann1973, okamoto1974,
 mestel1979, michel1990, sulkanen1990, beskin1993, beskin1998, mestel1999, contopoulos1999,
 ogura2003, goodwin2004, contopoulos2005, gruzinov2005, timokhin2006, petrova2013}.
 With thirty percent of the neutron star mass composed of free electrons and ions,
 the neutron star is regarded as a perfect conductor,
 and probably as a superconductor
 with the stellar magnetic field $\vec B$ trapped axially.
 As it rotates under this magnetic field,
 according to the unipolar induction,
 charges are driven to the polar and equatorial regions according to their signs,
 thus setting up a potential distribution on the pulsar surface.
 This potential distribution is uploaded to the magnetosphere
 through the magnetic field lines
 together with the charges, continuously pumped by the pulsar rotation,
 to keep the surface charge in steady state.
 Due to the electron mobility in shorting out
 any potential fluctuations along a field line,
 these field lines are equipotential lines.
 This generates a negatively charged plasma at high latitudes
 and a positively charged plasma at low latitudes.
 Being a charge separated plasma
 \citep{goldreich1969},
 the plasma flow thus generates a current flow, Eq.~\ref{eqno29}.
 The space around the pulsar is divided
 into the near zone with charge separated plasmas,
 the transition zone with quasi-neutral plasma,
 and the interstellar zone with interstellar plasma
 characterized by the reference floating potential of interstellar space.
 The current flow along the open field lines
 is driven by the potential difference
 between the pulsar surface and the interstellar floating potential.
 The open field line of the pulsar magnetosphere
 at the floating potential has no current flow.
 Those open field lines with higher latitude, thus lower potential,
 have current inflow (current from interstellar zone to the pulsar),
 and those with lower latitude have current outflow.
 As for the structure of the magnetosphere,
 it is generally regarded as dipole-like
 by similarity to a normal magnetic star.
 The magnetic field is considered as rigidly anchored on the pulsar surface
 and thus rotates with the pulsar, dragging the magnetosphere plasma along
 on behalf of the magnetic pressure and the plasma ram pressure.
 The plasma velocity increases linearly as the radius, in cylindrical coordinates,
 until it reaches the speed of light.

We note that a rigidly anchored magnetic field that rotates with the pulsar
 does not generate unipolar induction
 because the relative velocity between the rotating field
 and the rotating pulsar is null.
 This can be resolved by simply considering a stationary magnetic field
 under which the pulsar rotates.
 As a result, charges are rightfully driven to the polar and equatorial regions
 according to their signs
 setting up a potential distribution on the pulsar surface.
 After uploading this potential distribution to the magnetosphere
 through the magnetic field lines together with the charges,
 it generates a transverse electric field $\vec E$ across the field lines.
 With the electrons gyrating along the curvature
 of the intense pulsar magnetic field lines at the cyclotron frequency,
 they radiate high energy photons to produce electron-positron pairs.
 With the magnetosphere filled by an electron-positron plasma,
 together with the charges transported from the pulsar surface
 plus the presence of the magnetic $\vec B$ and electric $\vec E$ fields,
 they generate a unique magnetohydrodynamic (MHD) plasma for the pulsar magnetosphere.
 This unique MHD plasma differs from the normal quasi-neutral MHD plasma of magnetic stars
 and laboratory fusion devices on the degree of charge inequality
 between positive $n_{+}$ and negative $n_{-}$ fluid components.
 For the quasi-neutral laboratory MHD plasma,
 this degree of deviation $\Delta n/n=(n_{+}-n_{-})/n$ is insignificant,
 say $10^{-15}$ for example,
 and the Coulomb force is negligible in equilibrium configurations
 with respect to the magnetic force.
 The magnetosphere of a magnetic star is therefore described
 by the force-free magnetic field $\vec J\times\vec B=0$ only.
 For the pulsar MHD plasma, this degree of deviation is much higher
 due to the negative transported charges at high latitudes
 and positive transported charges at low latitudes
 from the pulsar surface to the electron-positron plasma.
 But it is still small, say $10^{-5}$ for example.
 However it reaches to the level
 that the Coulomb force has to be taken into consideration together with the magnetic force
 to describe pulsar magnetosphere equilibrium, Eq.~\ref{eqno1}.
 As for the center of mass plasma velocity,
 under the MHD Ohm law with infinite conductivity $\sigma$

\begin{eqnarray}
\nonumber
\vec E + \vec v\times\vec B\,
 =\,{1\over\sigma}\vec J\,
 =\,0\,\,\,,
\\
\nonumber
\vec v\,=\vec v_{E}\,
 =\,{1\over B^{2}}(\vec E\times\vec B)\,\,\,,
\end{eqnarray}

\noindent the plasma in the magnetosphere is set in motion
 at the angular velocity of the pulsar
 $\vec v_{E}=\vec r\times\vec\Omega_{0}$
 with respect to the stationary $\vec E$ and $\vec B$ fields
 by the $\vec E\times\vec B$ drift
 \citep{thompson1962, schmidt1966, jackson1975}.
 In a MHD plasma, we note that the Ohm law determines plasma flow,
 not plasma current.
 This offers an alternative description
 to the plasma being dragged by the rotating field.

The current pulsar model assumes that radiation is emitted continuously
 along the magnetic axis of a neutron star.
 With an offset between the magnetic axis and the rotational axis,
 this radiation could be detected by a distant observer at periods of the rotation.
 Nevertheless, since the first pulsar detection,
 this period has been reduced from seconds to milliseconds currently,
 down by three orders of magnitude.
 Considering a neutron star with radius $r_{ns}=10\,Km=10^{4}\,m$
 and rotation period $T=10^{-3}\,s$,
 the velocity of a given point on the equator
 reaches the relativistic velocity of $v=2\pi r_{ns}/T=6.3\times 10^{7}\,m/s$
 and the light cylinder is less than $50\,Km$ from the stellar center.
 Considering the pulsar magnetosphere being bounded
 by the light cylinder singularity, Eq.~\ref{eqno8},
 the rotation interpretation of the pulsar period is reaching to its limit.
 Here we attempt to describe the pulsar period
 as a result of its magnetosphere dynamics
 instead of its rotation period.

\newpage
\section{Force-Free Pulsar Magnetosphere}

Under the Coulomb force and the magnetic force,
 the pulsar magnetosphere is described by

\begin{eqnarray}
\label{eqno1}
\rho_{q}\vec E + \vec J\times\vec B\,=\,0\,\,\,,
\\
\nonumber
\rho_{q}\,=\,\pm e\Delta n\,\,\,,
\end{eqnarray}

\noindent where $\rho_{q}=\epsilon_{0}\nabla\cdot\vec E$
 measures the net deviation between electron and ion charges.
 Considering the electron-positron plasma $n_{ep}$ as quasi-neutral,
 this deviation comes from the latitude dependent unipolar pumped uploaded charges $\Delta n$.
 In cylindrical coordinates $(r,\phi,z)$,
 the axisymmetric magnetic field and current density can be represented as

\begin{eqnarray}
\label{eqno2}
\vec B\,=\,A_{0}\nabla\Psi\times\nabla\phi\,
 =\,A_{0}{1\over r}
 \left(-{\partial\Psi\over\partial z}, 0, +{\partial\Psi\over\partial r}\right)\,\,\,,
\\
\label{eqno3}
\mu_{0}\vec J\,=\,\nabla\times\vec B\,
 =\,A_{0}{1\over r}
 \left({0, -\nabla^{2}\Psi+{2\over r}{\partial\Psi\over\partial r}, 0}\right)\,\,\,,
\end{eqnarray}

\noindent where $A_{0}$ carries the dimension
 of the poloidal magnetic flux of the magnetosphere
 such that $\Psi$ is a dimensionless flux function.
 From the magnetic field line equation,
 the poloidal field lines are given by the contours of $\Psi$

\begin{eqnarray}
\label{eqno4}
{dr\over B_{r}}\,=\,{dz\over B_{z}}\,\,\,,
\\
\label{eqno5}
\Psi(r,z)\,=\,C\,\,\,.
\end{eqnarray}

Under the intensity of a given pulsar magnetic field line,
 the cyclotron frequency is many orders of magnitude
 above the electron collision frequency.
 Any potential fluctuation along a field line
 would be shorted out due to the electron mobility,
 thus $\vec E_{\parallel}=0$ along that magnetic field line.
 With the magnetic field $\vec B$ represented by Eq.~\ref{eqno2},
 the transverse electric field in the magnetosphere,
 generated by uploading the pulsar surface potential,
 can be expressed through the MHD Ohm law under infinity conductivity
 $\vec E=-\vec v\times\vec B=-(\vec r\times\vec\Omega)\times\vec B$ as

\begin{eqnarray}
\label{eqno6}
\vec E_{\perp}\,=\,-V_{0}\nabla\Psi\,
 =\,-A_{0}\Omega(\vec r)\nabla\Psi\,\,\,,
\end{eqnarray}

\noindent where $V_{0}$ is the equator-pole voltage drop
 on the pulsar surface by unipolar induction,
 $\Psi$ is the corresponding surface potential label plus an additive constant,
 and $\Omega(\vec r)$ is the plasma angular velocity of rotation
 in the stationary magnetosphere.
 With the Faraday law of induction

\begin{eqnarray}
\nonumber
\nabla\times\vec E_{\perp}\,=\,-A_{0}\nabla\Omega\times\nabla\Psi\,
 =\,-{\partial\vec B\over\partial t}\,\,\,,
\end{eqnarray}

\noindent we have in steady state that the angular velocity is a function of $\Psi$

\begin{eqnarray}
\label{eqno7}
\Omega(\vec r)\,=\,\Omega(\Psi(\vec r))\,\,\,.
\end{eqnarray}

\noindent Considering a rigid rotor $\Omega(\vec r)=\Omega_{0}$,
 the poloidal components of Eq.~\ref{eqno1} yield the pulsar equation

\begin{eqnarray}
\label{eqno8}
\left[1-({r\over r_{L}})^{2}\right]\nabla^{2}\Psi
 -{2\over r}{\partial\Psi\over\partial r}\,
 =\,0\,\,\,,
\end{eqnarray}

\noindent where $r_{L}=c/\Omega_{0}$ is the light cylinder (LC) radius,
 with the $\vec E\times\vec B$ plasma drift reaching the relativistic limit,
 and $\Omega_{0}$ is the angular velocity of the neutron star.
 In the rotating field line description,
 the location of this LC radius can be displaced within the near zone
 by choosing a soft plasma rotation function, instead of a rigid rotor.
 Physically, this soft rotation function represents dissipative effects of the plasma.
 Should we consider a differential rotor with $\Omega(\Psi(\vec r))$,
 the pulsar equation would read

 \begin{eqnarray}
\nonumber
\left[1-{r^2\over (c/\Omega(\Psi))^2}\right]\nabla^{2}\Psi
 -{r^2\over c^2}\Omega(\Psi){\partial\Omega(\Psi)\over\partial\Psi}(\nabla\Psi)^2
 -{2\over r}{\partial\Psi\over\partial r}\,
 =\,0\,\,\,,
\end{eqnarray}

\noindent which would be more appropriate to model a soft rotation function.

\newpage
\section{Singular Pulsar Equation}

The steady state magnetosphere with $(B_{r},B_{\phi},B_{z})$
 has been solved numerically notably by
 \citet{contopoulos1999},
 \citet{ogura2003},
 \citet{gruzinov2005},
 \citet{contopoulos2005}
 where the open and closed regions rotate differentially, and
 \citet{timokhin2006}
 where the braking index is evaluated
 in terms of the magnetosphere dynamics.
 Analytic solutions have been also developed by many authors
 \citep{scharlemann1973, okamoto1974, sulkanen1990}
 within LC and then continued beyond LC
 with physical variables matched across LC.
 In the published literature, it is important to note
 that \citet{sulkanen1990} have pioneered a pulsar magnetosphere
 having a polar jet.
 In these studies,
 the LC location has been treated as a regular point
 with physical variables continued smoothly beyond LC to the interstellar space.
 Nevertheless, because of the coefficient of the highest derivative
 becomes null at LC with $r/r_{L}=1$,
 Eq.~\ref{eqno8} is a singular equation
 \citep{morse1953}
 whose domain is divided by LC and has to be solved accordingly.
 Here, we attempt to construct an open and a closed magnetosphere
 of Eq.~\ref{eqno8} bounded by LC
 that can be matched to the void interstellar space across LC.
 With this objective, we consider a poloidal magnetic field configuration
 of Eq.~\ref{eqno2} with $B_{\phi}=0$.
 The presence of a toroidal field in the magnetosphere
 would generate a finite $B_{\phi}$ on LC
 and would be continued across LC,
 making the match to the void interstellar space impossible.
 To be consistent to the neutron star as a superconductor,
 we consider an axial stellar magnetic field at the center
 to operate the unipolar induction.

To solve for Eq.~\ref{eqno8}, we introduce a separation function $f(\Psi)$,
 analogous to the method of separation of variables
 through a separation constant.
 Denoting $\xi=r/r_{L}$ as the normalized radial coordinate, we get

\begin{eqnarray}
\label{eqno9}
(1-\xi^{2})\nabla^{2}\Psi\,
 =\,{1\over r_{L}^{2}}{2\over\xi}{\partial\Psi\over\partial\xi}\,
 =\,f(\Psi)\,\,\,,
\end{eqnarray}

\noindent where we have required both sides of the equality
 be equal to $f(\Psi)$.
 By choosing

\begin{eqnarray}
\label{eqno10}
f(\Psi)\,=\,k^{2}\Psi\,\,\,,
\end{eqnarray}

\noindent the pulsar equation renders

\begin{eqnarray}
\label{eqno11}
(1-\xi^{2})
 \left({\partial^{2}\Psi\over\partial\xi^{2}}
 +{1\over\xi}{\partial\Psi\over\partial\xi}
 +{\partial^{2}\Psi\over\partial\varsigma^{2}}\right)\,
 =\,(kr_{L})^{2}\Psi\,\,\,,
\\
\label{eqno12}
{2\over\xi}{\partial\Psi\over\partial\xi}\,
 =\,(kr_{L})^{2}\Psi\,\,\,,
\end{eqnarray}

\noindent where $\varsigma=z/r_{L}$ is the normalized axial coordinate.
 Let us write $\Psi(\xi,\varsigma)=R(\xi)Z(\varsigma)$ by separation of variables,
 and Eq.~\ref{eqno11} gives

\begin{eqnarray}
\label{eqno13}
{\partial^{2}Z\over\partial\varsigma^{2}}
 \pm m^{2}Z\,
 =\,0\,\,\,,
\\
\label{eqno14}
(1-\xi^{2})
 \left({\partial^{2}R\over\partial\xi^{2}}
 +{1\over\xi}{\partial R\over\partial\xi}
 \mp m^{2}R\right)\,
 =\,(kr_{L})^{2}R\,\,\,,
\end{eqnarray}

\noindent where $m^{2}$ is the separation constant.
 As for Eq.~\ref{eqno12}, we get

\begin{eqnarray}
\label{eqno15}
{2\over\xi}{\partial R\over\partial\xi}\,
 =\,(kr_{L})^{2}R\,\,\,.
\end{eqnarray}

\noindent Making use of Eq.~\ref{eqno15} on Eq.~\ref{eqno14} we get

\begin{eqnarray}
\label{eqno16}
  (1-\xi^{2}){\partial^{2}R\over\partial\xi^{2}}
 +(1-\xi^{2})\left[{1\over 2}(kr_{L})^{2}\mp m^{2}\right]R
 -(kr_{L})^{2}R\,
 =\,0\,\,\,.
\end{eqnarray}

\noindent Casting the above equation in the form

\begin{eqnarray}
\nonumber
  (1-\xi^{2}){\partial^{2}R\over\partial\xi^{2}}
 +(1-\xi^{2})\alpha_{\mp}R
 -\beta R\,
 =\,0\,\,\,,
\\
\label{eqno17}
\left({\partial^{2}R\over\partial\xi^{2}}
 +\alpha_{\mp}R\right)\,
 =\,{\beta\over (1-\xi^{2})}R\,\,\,,
\end{eqnarray}

\noindent we note that the LC singularity at $\xi=1$,
 which divides the $\xi$ domain in two parts with $\xi<1$ and $\xi>1$,
 is preserved in the present form of separation through Eq.~\ref{eqno9}.
 To understand qualitatively the behavior of this equation,
 we recall that the left side is the sinusoidal operator
 and the right side is the LC singularity operator.
 Over the range of $\xi$,
 the solution $R(\xi)$ is the result of the competition between these two operators
 weighted by the two coefficients $\alpha_{\mp}$ and $\beta$.
 This equation together with Eq.~\ref{eqno13}
 describe the structure of the pulsar magnetosphere.

To solve for Eq.~\ref{eqno13}, we first consider the upper sign to get

\begin{eqnarray}
\label{eqno18}
Z_{a}(\varsigma)\,
 =\,\cos{m\varsigma}\,\,\,,
\\
\label{eqno19}
Z_{b}(\varsigma)\,
 =\,\sin{m\varsigma}\,\,\,.
\end{eqnarray}

\noindent The domain of these solutions is bounded between $(0,\,\pi)$.
 As for the lower sign, we have

\begin{eqnarray}
\label{eqno20}
Z_{a}(\varsigma)\,
 =\,e^{+m\varsigma}\,\,\,,
\\
\label{eqno21}
Z_{b}(\varsigma)\,
 =\,e^{-m\varsigma}\,\,\,.
\end{eqnarray}

\noindent This solution covers an unbounded domain of $\varsigma$.
 To solve for Eq.~\ref{eqno17} numerically for the singular $R(\xi)$,
 we need to define the boundary condition at $\xi=0$.
 To be consistent to the idea that the neutron star being a superconductor,
 we consider the stellar magnetic field being axial
 in contrast to a dipole-like source field of a magnetic star.
 With the magnetic field components given by

\begin{eqnarray}
\nonumber
B_{r}\,
 =\,{1\over r^{2}_{L}}A_{0}{1\over\xi}R(\xi){\partial Z(\varsigma)\over\partial\varsigma}\,\,\,,
\\
\nonumber
B_{z}\,
 =\,{1\over r^{2}_{L}}A_{0}{1\over\xi}{\partial R(\xi)\over\partial\xi}Z(\varsigma)\,\,\,,
\end{eqnarray}

\noindent we require $B_{r}=0$ at the origin $\xi=\varsigma=0$.
 We therefore have either $R(\xi)/\xi=0$
 or $\partial Z(\varsigma)/\partial\varsigma=0$ for $B_{r}=0$.
 For $R(\xi)$, we therefore have the asymptotic form of $R(\xi)=\xi^{2}$
 in the vicinity of $\xi=0$.
 Since we consider either the upper sign with $Z_{a}(\varsigma)$ of Eq.~\ref{eqno18},
 or the lower sign with $Z_{b}(\varsigma)$ of Eq.~\ref{eqno21},
 both with finite value at $\varsigma=0$,
 to construct the magnetosphere,
 the boundary condition rests on $R(\xi)=\xi^{2}$ entirely.
 We therefore require
 $R(\xi)=\xi_{0}=0$ and $\partial R(\xi)/\partial\xi=2\xi_{1}$ at the origin
 for the first two grid points $\xi_{0}=0\delta\xi$ and $\xi_{1}=1\delta\xi$
 with $\delta\xi$ as the grid size.

\newpage
\section{LC Guided Magnetospheres}

We note that Eq.~\ref{eqno17} is normalized in dimensionless form.
 It is therefore important to assess the order of magnitude of the coefficients.
 First, we note that $k$ in the separating function $f(\Psi)$
 is the inverse scale length of $\Psi$
 as indicated by Eq.~\ref{eqno9} and Eq.~\ref{eqno10}.
 Thus $k$ measures the transverse field gradient in the magnetosphere.
 Second, the $m^{2}$ term in $\alpha_{\mp}$ is of the order of unity,
 and it is negligible comparing to the $(kr_{L})^2$ term.
 In this pulsar magnetosphere,  there are two characteristic distances.
 The first one is the neutron star radius $r_{ns}$
 and the second one is the light cylinder radius $r_{L}$.
 The magnetic field from the volume of $r_{ns}$
 would expand into the space bounded by $r_{L}$.
 As an example, to assess the coefficients $\alpha_{\mp}$ and $\beta$,
 we consider a neutron star with $r_{ns}=10\,Km$ and rotation period $T_{ns}=1\,s$,
 which gives $r_{L}=5\times 10^{4}\,Km$.
 Should we consider the magnetic field scale length be $10^{3}\,Km$,
 we would have $(kr_{L})^{2}=(50)^{2}=2.5\times 10^{3}$,
 which would be the order of magnitude of $\alpha_{\mp}$ and $\beta$.
 Further, we note that the LC singularity of Eq.~\ref{eqno17}
 becomes dominant only at the vicinity of $\xi=1$.
 Should $\beta=0$, Eq.~\ref{eqno17} would describe a sinusoidal solution.
 Should $\alpha_{\mp}=0$, it would describe a singular solution.
 Therefore, the ratio $\alpha_{\mp}/\beta$ is an important parameter
 in balancing the sinusoidal and singular behaviors.

Taking the corresponding lower sign of $\alpha_{\mp}$ in Eq.~\ref{eqno17}
 and considering $\alpha_{+}=30$ and $\beta=30$ with  $\alpha_{+}/\beta=1$ ,
 the function $R(\xi)$ is plotted in Fig.1
 showing an increasing function towards LC for an open configuration.
 Considering the lower sign of Eq.~\ref{eqno13} with $Z_{b}(\varsigma)$ of Eq.~\ref{eqno21},
 the corresponding set of poloidal field lines

\begin{eqnarray}
\label{eqno22}
\Psi(\xi,\varsigma)\,
 =\,R(\xi)Z_{b}(\varsigma)\,
 =\,R(\xi)e^{-m\varsigma}
 =\,C\,\,\,,
\end{eqnarray}

\noindent is shown in Fig.2 with $m=1$ for different contour values $C$
 originating from the neutron star
 with a hypothetical normalized radius $\xi_{ns}=0.1$.
 We note that the field lines approach LC at $\xi=1$
 only as $\varsigma\rightarrow\infty$.
 As a result, the field lines only reach LC asymptotically.
 Let us now examine the asymptotic limit of the field components.
 As the field lines approach LC at $\xi=1$ as $\varsigma\rightarrow\infty$,
 the radial component $B_{r}$ becomes null.
 This is evident from the expression of $B_{r}$,
 since $Z(\varsigma)$ grows exponentially in the denominator
 because of the negative power of $Z_{b}(\varsigma)$.
 This dependence overrides the power law growth of $R(\xi)$.
 As for the axial component $B_{z}$,
 it becomes null asymptotically as well
 because $\partial R(\xi)/\partial\xi\rightarrow 0$.
 Thus the field is null on LC and is also null outside LC by continuity.
 The magnetic field fills LC only asymptotically.
 For small values of $\varsigma$,
 the magnetic field occupies only the space
 near the pulsar with $\xi<1$.
 This solution describes a jetted magnetosphere within LC
 which smoothly matches the void interstellar space across LC.
 We note that astrophysical jets
 are usually accompanied by equatorial accretion disk
 that provides the primary angular momentum
 to be collimated onto the jet structure
 making it stable dynamically.
 For the present pulsar jet,
 this angular momentum comes from the
 $\vec v_{E}=(\vec E\times\vec B)/B^{2}$ drift velocity of the plasma,
 where the electric field $\vec E$ is originated
 by uploading the unipolar driven potential distribution
 on the pulsar surface to the magnetic field lines.
 Thus the primary source of the jet angular momentum
 is the neutron star rotation.
 This jetted magnetosphere is in line with the work of
 \citet{sulkanen1990}
 who proposed a pulsar magnetosphere having a polar jet
 described by the hypergeometric series.

\begin{figure}[b!]
\includegraphics[scale = 0.5]{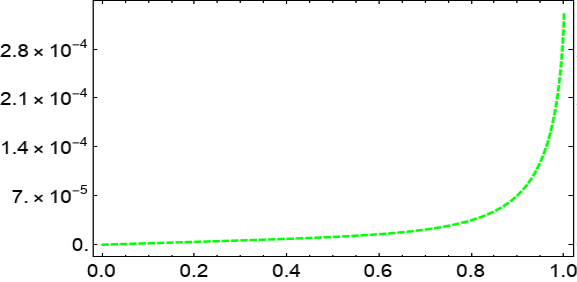}
\caption{\footnotesize The radial function $R(\xi)$ with $\alpha_{+}=30$ and $\beta=30$
 is shown to have an increasing profile towards LC
 with respect to the normalized radial distance $\xi$.}
\label{fig.1}
\normalsize
\end{figure}

\begin{figure}[b!]
\includegraphics[scale = 0.5]{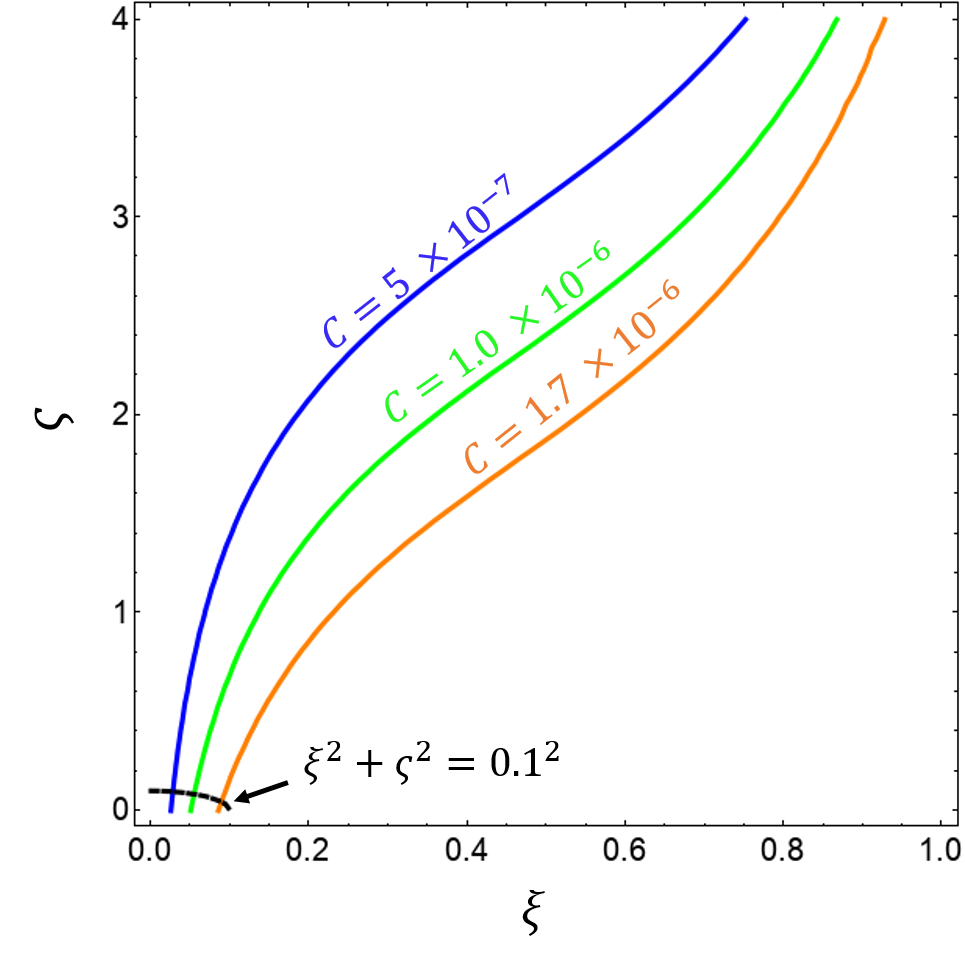}
\caption{\footnotesize A set of poloidal field lines with $m=1$
 originating from the neutron star with $\Psi(\xi,\varsigma)=C$
 for an open magnetosphere is shown with $\xi_{ns}=0.1$.}
\label{fig.2}
\normalsize
\end{figure}

\begin{figure}[b!]
\includegraphics[scale = 0.5]{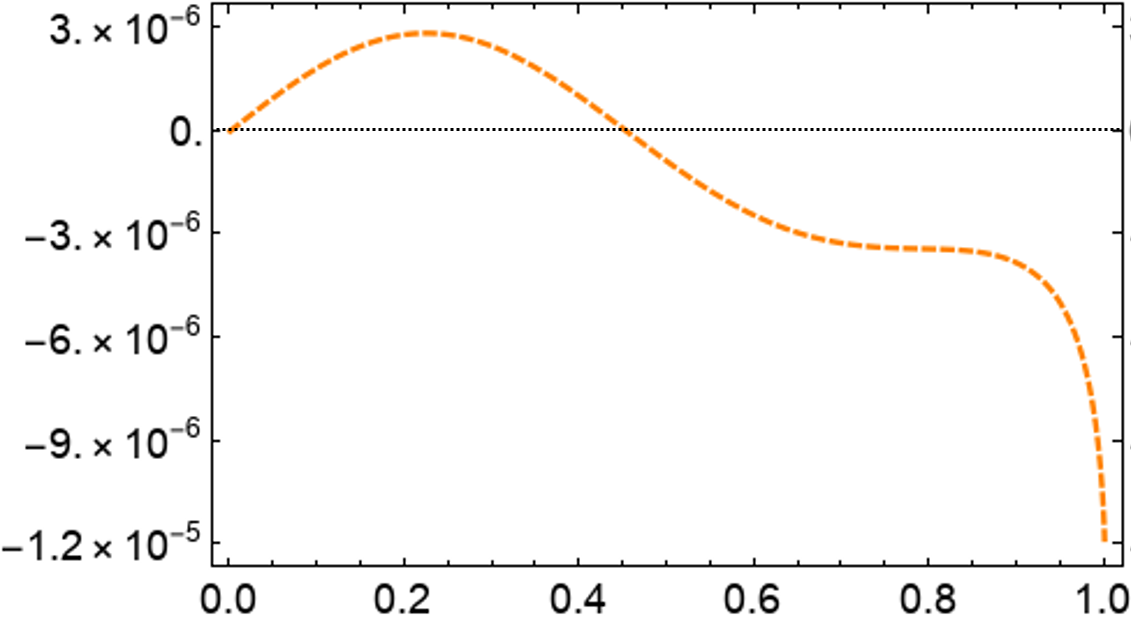}
\caption{\footnotesize The radial function $R(\xi)$ with $\alpha_{+}=80$ and $\beta=30$
 is shown to have an oscillatory profile before reaching LC
 with respect to the normalized radial distance $\xi$.}
\label{fig.3}
\normalsize
\end{figure}

\begin{figure}[b!]
\includegraphics[scale = 0.5]{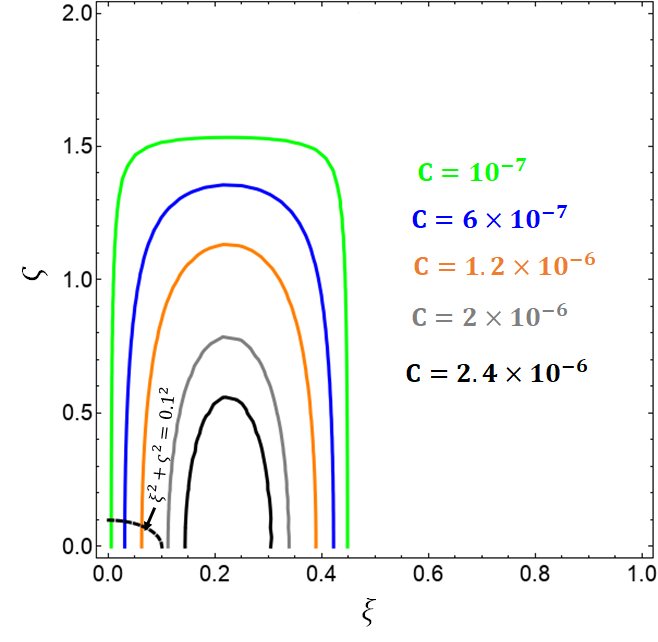}
\caption{\footnotesize A set of poloidal field lines with $m=1$
 originating from the neutron star with $\Psi(\xi,\varsigma)=C$
 for a closed magnetosphere is shown with $\xi_{ns}=0.1$.}
\label{fig.4}
\normalsize
\end{figure}

Now, let us consider the upper sign of Eq.~\ref{eqno13}
 with $Z_{a}(\varsigma)$ of Eq.~\ref{eqno18}.
 With either $\partial Z_{a}(\varsigma)/\partial\varsigma=0$ at $\varsigma=0$
 or the boundary condition $R(\xi)=\xi^{2}$ around $\xi=0$,
 we have $B_{r}=0$ at the center.
 We therefore have an axial magnetic field on the axis for unipolar induction.
 The corresponding upper sign of Eq.~\ref{eqno17} for $R(\xi)$
 with $\alpha_{-}=80$ and $\beta=30$ and $\alpha_{-}/\beta=8/3$ is plotted in Fig.3
 showing the competition between the sinusoidal and the singular terms.
 This figure shows a local maximum $\xi^{*}_{0}$
 bounded by a null point $\xi^{*}_{1}$,
 which is a separatrix for the closed configuration,
 before reaching LC asymptotically.
 A corresponding set of poloidal field lines

\begin{eqnarray}
\label{eqno23}
\Psi(\xi,\varsigma)\,
 =\,R(\xi)Z_{a}(\varsigma)\,
 =\,R(\xi)\cos m\varsigma\,
 =\,C\,\,\,,
\end{eqnarray}

\noindent is shown in Fig.4 with $m=1$ indicating a closed magnetosphere
within the null point around $\xi=0.45$ where $R(\xi)=0$.
To do this closed magnetosphere,
we need only $\cos m\varsigma$ over $0<\varsigma<\pi/2$.
Naturally, the highly exaggerated dimension with $\xi_{ns}=0.1$
of the open and closed magnetospheres is for illustration purposes.
Increasing the contour value $C$,
the closed poloidal field line would be detached from the pulsar surface.
Further increasing $C$, more magnetic contours would be generated within earlier ones
until reaching the magnetic axis around $\xi=0.23$ where $dR(\xi)/d\xi=0$.
We remark that we have constructed the open and closed magnetospheres
bounded by LC using an axial magnetic field on the $z$ axis
as the boundary condition.
Should we choose a dipole source with infinitely dense field lines
at the stellar center,
we would get the familiar closed configuration up to LC.

\newpage
\section{Revised Goldreich Julian Condition - Magnetosphere Dynamics}

We have solved the pulsar equation, Eq.~\ref{eqno8},
 for two equilibrium configurations.
 We remark that the pulsar equation is just the result of the equation of motion
 with the inertial term neglected for force-free field equilibrium
 coupled with the Maxwell equations.
 This equation by itself is model independent,
 and it depends on how the other physical variables are defined,
 such as the charge density, current density, and others.
 We note that the Ampere law provides the curvature of $\vec B$
 in the presence of a current density $\vec J$.
 In a conductor, current density can be generated
 in the presence of an electromotive force,
 or by rotating it under an external magnetic field.
 In a magnetically confined MHD plasma,
 current can be generated by the guiding center drifts
 of the electron and ion species
 due to the magnetic curvature and gradient
 \citep{thompson1962, schmidt1966, jackson1975}.
 We recall that in a uniform magnetic field,
 charge particles follow a circular gyro motion
 along the magnetic field lines
 according to the equation of motion.
 In the presence of a force transverse to the magnetic field,
 such as a centrifugal force due to curvature,
 and an inhomogeneous magnetic field, etc.,
 the circular gyro motion will be transformed
 into a slowly displacing cycloid across the field lines,
 such that the gyro guiding center
 presents a transverse drift.
 These drifts are given by

\begin{eqnarray}
\label{eqno24}
\vec v_{Rc}\,
 =\,+{2\epsilon_{\parallel}\over q}
 {1\over B^{2}}
 ({\vec R_{c}\over R_{c}^{2}}\times\vec B)\,\,\,,
\\
\label{eqno25}
\vec v_{\nabla B}\,
 =\,-{\epsilon_{\perp}\over q}
 {1\over B^{2}}
 ({\nabla B\over B}\times\vec B)\,\,\,,
\end{eqnarray}

\noindent where $\epsilon_{\parallel}=mv_{\parallel}^{2}/2$
 and $\epsilon_{\perp}=mv_{\perp}^{2}/2$
 are the charge particle energies (temperature)
 parallel and perpendicular to the magnetic field.
 Therefore, these drifts are mass dependent
 as well as field dependent.
 The direction of these drifts depends on the sign of charge q,
 thus electrons and ions go in opposite direction generating a current,

\begin{eqnarray}
\label{eqno26}
\vec J\,
 =\,\rho_{q}^{*}\vec v_{current}\,
 =\,\rho_{q}^{*}(\vec v_{Rc}+\vec v_{\nabla B})\,
 =\,en(\vec v_{Rc}+\vec v_{\nabla B})\,\,\,,
 \end{eqnarray}

\noindent where $\rho_{q}^{*}=en$ is the current generating charge density
 with all charged particles taking part.
 In the pulsar magnetosphere,
 the dominant part is the electron-positron plasma $n_{ep}$
 plus the uploaded charges.
 Furthermore $\vec v_{current}$ is the current generating flow.
 As a result, the plasma drifts generate a current density $\vec J$
 that feeds back to the magnetic field $\vec B$ through the Ampere law.
 We can now evaluate the Coulomb force space charge density

\begin{eqnarray}
\nonumber
\rho_{q}\,=\,\epsilon_{0}\nabla\cdot\vec E\,
 =\,-\epsilon_{0}\nabla\cdot(\vec v_{E}\times\vec B)\,
 =\,\epsilon_{0}(\vec v_{E}\cdot\mu_{0}\vec J
 -\vec B\cdot\nabla\times\vec v_{E})\,\,\,,
\end{eqnarray}

\noindent which can be written as

\begin{eqnarray}
\label{eqno27}
[\rho_{q}-\epsilon_{0}\mu_{0}\vec v_{E}\cdot\vec J]\,
 =\,-\epsilon_{0}\vec B\cdot\nabla\times\vec v_{E}\,\,\,.
\end{eqnarray}

\noindent By using Eq.~\ref{eqno26}, we then have

\begin{eqnarray}
\label{eqno28}
[\rho_{q}-\epsilon_{0}\mu_{0}\rho_{q}^{*}\vec v_{E}\cdot\vec v_{current}]\,
 =\,-\epsilon_{0}\vec B\cdot\nabla\times\vec v_{E}\,
 =\,-2\epsilon_{0}\vec B\cdot\vec\Omega_{0}\,
 =\,-2\epsilon_{0}B_{z}\Omega_{0}\,\,\,,
\end{eqnarray}

\noindent where the last equality is evaluated through $\vec v_{E}=r\Omega_{0}\hat\phi$.

Without considering the electron-positron plasma,
 \citet{goldreich1969} had derived a condition
 for the space charge density (their Eq.8)
 that read

\begin{eqnarray}
\label{eqno29}
\vec J\,=\,\rho_{q}^{*}\vec v_{current}\,
=\,\rho_{q}^{*}\vec v\,=\,\pm en\vec v\,\,\,,
\\
\label{eqno30}
(1-\epsilon_{0}\mu_{0}v^{2})\rho_{q}^{*}\,
 =\,-2\epsilon_{0}\vec B\cdot\vec\Omega_{0}\,
 =\,-2\epsilon_{0}B_{z}\Omega_{0}\,\,\,.
\end{eqnarray}

\noindent They had denoted $\vec v$ as the plasma flow velocity, equivalent to $\vec v_{E}$,
 but this plasma flow velocity was also the current generating velocity, as in Eq.~\ref{eqno29}.
 Current flow was generated by plasma flow
 because the positive and negative plasmas were spatially separated in their model.
 The closed magnetosphere in the equatorial region
 was generated by the positive plasma with $\rho_{q}^{*}>0$.
 As for the open magnetosphere,
 it was generated by the open field lines of the negative plasma in the polar region
 plus part of the positive plasma in the intermediate region.
 The direction of the poloidal current flow along these open field lines
 was driven by the potential difference
 between the pulsar surface and the floating potential in the interstellar space,
 with a null line (null current flow) separating the inflow and outflow currents.
 In this charge separated model,
 all the charges took part in generating current through the plasma flow
 rendering $\vec J=\pm en\vec v=\rho_{q}^{*}\vec v$.
 As a result, the charge density of the Coulomb force $\rho_{q}\vec E=\pm en\vec E$
 and the charge density of current generation
 $\vec J=\rho_{q}^{*}\vec v=\pm en\vec v$ were the same,
 $\rho_{q}=\rho_{q}^{*}=\pm en$,
 contrary to the $\rho_{q}$ of Eq.~\ref{eqno1}.
 Applying Eq.~\ref{eqno30} to the closed equatorial magnetosphere
 with $\rho_{q}=\rho_{q}^{*}=+en>0$ and
 with the bracket multiplying $\rho_{q}^{*}$ on the left side always positive,
 this celebrated GJ condition indicated $B_{z}<0$ always.
 Thus the closed positively charged equatorial field lines
 remained closed and hence the term ¨dead zone¨,
 in contrast to the open field lines.
 For this reason, Eq.~\ref{eqno30} described the stability
 of the positively charged closed pulsar magnetosphere.

Currently, although the magnetosphere plasma is regarded as a MHD plasma,
 the notion of a dead zone closed magnetosphere has persisted in MHD descriptions,
 and Eq.~\ref{eqno28} appears to be compatible to Eq.~\ref{eqno30} showing the dead zone.
 But it is not for two reasons.
 First, under the present description of a MHD plasma,
 with an electron-positron background plasma plus the uploaded unipolar pumped surface charges,
 the center of mass flow $\vec v_{E}$ of a MHD plasma does not generate current
 because the electrons and ions drift in the same direction,
 only the current generating flows of Eq.~\ref{eqno24} and Eq.~\ref{eqno25}
 are responsible for current.
 This is in contrast with the charge separated plasma model of Eq.~\ref{eqno29}
 where plasma current is generated by charged plasma flow with $\rho_{q}=\pm en$.
 Second, the charge density of the Coulomb force in Eq.~\ref{eqno1}
 is given by $\rho_{q}=\pm e\Delta n$.
 On the other hand, for the current density of Eq.~\ref{eqno26} generated by drifts,
 all charged particles take part,
 but predominantly the electron-positron plasma $n_{ep}$,
 with $\rho_{q}^{*}=\pm en=\pm en_{ep}$.
 For these two reasons, in terms of $\rho_{q}$ of Eq.~\ref{eqno1},
 the current density of Eq.~\ref{eqno26} becomes

\begin{eqnarray}
\label{eqno31}
\vec J\,=\,\rho_{q}^{*}\vec v_{current}\,
 =\,\rho_{q}{n\over\Delta n}\vec v_{current}\,\,\,.
\end{eqnarray}

\noindent Consequently, Eq.~\ref{eqno28} reads

\begin{eqnarray}
\label{eqno32}
[1-\epsilon_{0}\mu_{0}{n\over\Delta n}\vec v_{E}\cdot\vec v_{current}]\rho_{q}\,
 =\,-2\epsilon_{0}B_{z}\Omega_{0}\,\,\,.
\end{eqnarray}

\noindent Although the current generating drift velocity is quite slow,
 in the presence of the ${n/\Delta n}$ factor in Eq.~\ref{eqno32},
 the closed magnetosphere within LC could open up, with $B_{z}>0$,
 to a LC guided jet
 as the electron-positron plasma density in the magnetosphere builds up.
 Thus the pulsar magnetic field re-establishes the axial field configuration,
 and the whole cycle starts anew.

\newpage
\section{Conclusions}

Considering the neutron star as a superconductor
 rotating under a stationary axial magnetic field, instead of a dipole field,
 the unipolar induction continuously pumps charges
 to the pulsar surface according to their signs
 and sets up a potential distribution on the surface.
 These charges are uploaded to the magnetosphere along the magnetic field lines $\vec B$
 to keep the surface charges in steady state
 and to generate a transverse electric field $\vec E$ across the field lines
 setting the plasma in motion with the $\vec E\times\vec B$ drift.
 Together with the electron-positron plasma in the magnetosphere,
 they generate a unique MHD plasma for the pulsar magnetosphere
 which is characterized by the Coulomb force charge density $\rho_{q}=\pm e\Delta n$
 and the plasma current charge density $\rho_{q}^{*}=en$.
 Taking into consideration of the singularity,
 we have solved the pulsar equation
 for an open and a closed LC guided magnetosphere
 through a separation function with two parameters $(k^{2},\,m^{2})$.
 By choosing one set of parameters
 such that $R(\xi)$ is an increasing function asymptotically towards LC,
 we have constructed the polar jet configuration with $(\alpha_{+},\,\beta)$.
 The angular momentum of the toroidal $\vec E\times\vec B$ plasma drift
 collimates the jet making it stable over its lifetime.
 While with another set of parameters
 such that $R(\xi)$ goes through a maximum before approaching LC,
 we have also constructed the closed configuration with $(\alpha_{-},\,\beta)$.
 These two configurations reinforce the magnetosphere polar jet structure
 first pioneered by \citet{sulkanen1990}.
 By distinguishing between the Coulomb force charge density $\rho_{q}$
 and the plasma current charge density $\rho_{q}^{*}$,
 and between the center of mass flow $\vec v_{E}$
 and the current generating flow $\vec v_{current}$,
 the celebrated GJ condition for the closed magnetosphere is revised,
 which shows that the closed magnetosphere could open
 as the electron-positron plasma density builds up.
 Consequently, the pulsar magnetosphere alternates
 between the closed and open configurations due to magnetosphere dynamics,
 releasing the stored magnetosphere energy along the pulsar axis periodically.

\newpage
\centerline{\bf Funding}

This research did not receive funding.

\end{document}